\documentclass[aps,twocolumn,showpacs,preprintnumbers,nofootinbib,prd,superscriptaddress,10pt]{revtex4-1}

\makeatletter
\def\l@subsubsection#1#2{}
\def\l@subsubsubsection#1#2{}
\makeatother

\usepackage{comment}
\usepackage{graphicx,amssymb,amsmath,amsthm,amsfonts,epsfig,epsf,fixmath}
\usepackage[usenames]{color}
\usepackage{xcolor}
\usepackage{epstopdf}

\usepackage{aas_macros}
\usepackage{bm}
\usepackage{dcolumn}
\usepackage{lipsum}
\usepackage{latexsym}
\usepackage{rotating}
\usepackage{longtable}

\usepackage{enumerate}
\usepackage{tensor,multirow}
\usepackage{url}
\usepackage[linktocpage]{hyperref}

\begin{document}

%%%%%%%%%%%%%%%%%%%%%%%%%%%%%%%%%%%%%%%%%%%%%%%%%%%%%%%%%%%%%%%%%%%%%%%%%%%%%%%%%%%%%%%%%%%%%%%%%%%%%%%%%%%%%%%%%%%%%
\title{Plasma-photon interaction in curved spacetime.~I:\\Formalism and quasibound states around nonspinning black holes}
%%%%%%%%%%%%%%%%%%%%%%%%%%%%%%%%%%%%%%%%%%%%%%%%%%%%%%%%%%%%%%%%%%%%%%%%%%%%%%%%%%%%%%%%%%%%%%%%%%%%%%%%%%%%%%%%%%%%%
%\author{
%Enrico Cannizzaro$^1$,
%Andrea Caputo$^{2,3,4}$,
%Laura Sberna$^{5}$,
%Paolo Pani$^{1}$}

\author{Enrico Cannizzaro}
\affiliation{Dipartimento di Fisica, ``Sapienza'' Universit\`a di Roma \& Sezione INFN Roma1, Piazzale Aldo Moro 
5, 00185, Roma, Italy}
\author{Andrea Caputo}
\affiliation{School of Physics and Astronomy, Tel-Aviv University, Tel-Aviv 69978, Israel}
\affiliation{Department of Particle Physics and Astrophysics,
Weizmann Institute of Science, Rehovot 7610001, Israel}
\affiliation{Max-Planck-Institut f\"ur Physik (Werner-Heisenberg-Institut), F\"ohringer Ring 6, 80805 M\"unchen, Germany}
\author{Laura Sberna}
\affiliation{Max Planck Institute for Gravitational Physics (Albert Einstein Institute) Am Mu\"{u}hlenberg 1, 14476 Potsdam, Germany}
\affiliation{Perimeter Institute, 31 Caroline St N, Waterloo, ON N2L 2Y5, Canada}
\author{Paolo Pani}
\affiliation{Dipartimento di Fisica, ``Sapienza'' Universit\`a di Roma \& Sezione INFN Roma1, Piazzale Aldo Moro
5, 00185, Roma, Italy}

%----------------------------------------------------------------------------------------------------
\begin{abstract} 
We investigate the linear dynamics of an electromagnetic field propagating in curved spacetime in the presence of 
plasma. The dynamical equations are generically more involved and richer than the effective Proca equation adopted as a model in previous 
work. We discuss the general equations and focus on the case of a cold plasma in the background of a spherically symmetric black hole, showing that the system admits plasma-driven, quasibound electromagnetic states that are prone to become superradiantly unstable when the black hole rotates. The quasibound states are different from those of the Proca equation and have some similarities with the case of a massive scalar field, suggesting that the linear instability can be strongly suppressed compared to previous estimates. Our framework provides the first step toward a full understanding of the plasma-photon interactions around astrophysical black holes.
\end{abstract}
%----------------------------------------------------------------------------------------------------

\maketitle

%----------------------------------------------------------------------------------------------------
\section{Introduction}
%----------------------------------------------------------------------------------------------------
\subsection{Motivation}
The propagation of fundamental fields on black hole~(BH) spacetimes is a decades-long fascinating subject, which has received considerable attention in the last years for its relevance for tests of fundamental physics with BH observations~\cite{Barack:2018yly}. A particularly relevant case is the one of an electromagnetic~(EM) field propagating in a BH spacetime in the presence of plasma. 
The study of plasmas is crucial in many branches of physics and astrophysics, including the study of BHs, for at least two reasons: (i) astrophysical BHs can be surrounded by accretion disks due to the outward transfer of angular momentum of accreting matter~\cite{1973A&A....24..337S,Novikov:1973kta}; (ii) photons emitted by accreting BHs can interact with the plasma of the interstellar medium, potentially affecting their propagation at large distances.

It is common lore to assume that a photon propagating in a plasma is simply dressed with an effective mass, induced by the interaction with the background medium. A photon in a plasma is therefore commonly described with a Proca-like equation $\nabla_{\sigma} F^{\sigma\nu} = \omega_{\rm pl}^2 A^{\nu}$, where $F_{\sigma\nu}=\partial_\sigma A_\nu-\partial_\nu A_\sigma$ is the EM field strength and\footnote{Throughout this paper, we use $G=c=1$ and rationalized Heaviside units for the Maxwell equations. The metric signature is mostly positive.}
%%%
\begin{equation}
   \omega_{\rm pl}= (n e^2/m_e)^{1/2} \label{plasmafreq}
\end{equation}
%%%
is the plasma frequency of the background medium~\cite{Sitenko:1967}, with $n$ the particle density of electrons with charge $e$ and mass $m_e$. 
The Proca equation has been studied in detail on both Schwarzschild and Kerr metrics~\cite{Rosa:2011my,Pani:2012vp, Pani:2012bp,Baryakhtar:2017ngi,Cardoso:2018tly,Frolov:2018ezx,Dolan:2018dqv,Baumann:2019eav}. In the case of a rotating BH, the system can develop a superradiant instability, similar to the one exhibited by massive scalar fields (see Ref.~\cite{Brito:2015oca} for a review). This sows the seeds for striking observational signatures, which are particularly significant for primordial BHs dark matter~\cite{Pani:2013hpa} or in the presence of extra degrees of freedom such as axions~\cite{Blas:2020nbs}. This mechanism has been also advocated as a possible explanation for the origin of fast radio bursts~\cite{Conlon:2017hhi}. 

Given the interesting phenomenological implications of plasma-photon interactions around BHs, it is of crucial importance to understand first of all whether or not the Proca equation represents a good description of the dynamics of the system. 
The fact that the Proca equation can at most be an {\it approximation} is already clear from the counting of the degrees of freedom: a massive spin-1 field propagates three polarizations (two transverse modes and a longitudinal one), whereas photons in a cold plasma propagate only two transverse modes, since the putative longitudinal mode is electrostatic, see details below.

In this work we take the first step toward a detailed investigation of the propagation of EM waves in a plasma in curved spacetime, and show that even in the case of a Schwarzschild BH --~where of course superradiance cannot develop~-- the spectrum of the quasibound states deviates from the one of a simple Proca equation.
This will impact the plasma-driven superradiant instability, since the quasibound states around a static BH are those prone to become superradiantly unstable when the BH rotates~\cite{Pani:2012bp,Pani:2012vp,Brito:2015oca}.
Our results call for further investigation, and are likely pointing toward a strong suppression of the expected superradiant instability in rotating BH spacetimes. 

Other plasma effects have been pointed out recently and indicate that plasma may generically hinder our ability to perform strong-field gravity tests, for example by quenching the superradiance instability~\cite{Cardoso:2020nst, Blas:2020kaa}. The present work shows that the suppression of the instability will be even more severe (and already present at the linear level) when one accounts precisely for the presence of the plasma in the dynamical equations. 

In principle, the above problems could be studied in the framework of general-relativity-magneto-hydrodynamics~(GRMHD) (see, e.g., Refs.~\cite{Abramowicz:2011xu,Yuan:2014gma} for a review). However, GRMHD simulations typically focus on highly-magnetized plasmas in the absence of an electric field, and are also computationally expensive, which makes them unsuitable to follow the (potentially long) evolution of the microscopic EM field on different backgrounds. Indeed, to the best of our knowledge, a detailed understanding of the propagation of EM waves in a plasma in curved spacetime problem was not considered in previous work.

\subsection{Cold and hot plasma}
\label{sub:plasma}
A plasma is a gas composed of electrons and ions (ionized by heating or photoionization) which obey the classical Maxwell-Boltzmann rather than Fermi-Dirac or Bose-Einstein statistics~\cite{Sitenko:1967}. 
A plasma can be further classified in terms of its temperature. In a {\it cold plasma} the thermal velocity of the electrons is larger than that of ions, but still much smaller than the speed of an EM wave propagating in it, namely
%%%
\begin{equation}
   v_{\rm thermal} \equiv \sqrt{\frac{2 T_e}{m_e}} \ll \frac{ \omega}{k}\,, 
\end{equation}
%%%
where $\omega$ and $k$ are the frequency and wave number of the EM field, and $T_e$ is the temperature of the electrons. In this case the thermal pressure is negligible. In a cold plasma, the longitudinal modes are plasma oscillations which do not propagate and do not transport energy. The presence of this electrostatic mode already shows that the Proca equation cannot fully describe the interaction of photons with a cold plasma. 

On the other hand, in a {\it hot plasma}, the electron thermal velocity cannot be neglected and the electrons cannot be considered at rest with respect to propagating waves.  In this case the electrostatic modes are converted into propagating, energy-transporting, longitudinal modes called Langmuir waves~\cite{2017mcp..book.....T}. These modes propagate at the speed of sound in the plasma and the (nonrelativistic) equations resemble the Proca one in the case in which the electrons are ultrarelativistic~\cite{PhysRevD.45.525,PhysRevE.62.2989} (although this approximation is beyond the regime of validity of the original equations). 

In the following we will limit ourselves to the study of cold plasma, since the latter provides a good description of accretion disks around BHs~\cite{1974ApJ...191..499P, 1973A&A....24..337S}. Indeed, in the inner region of a typical accretion disk the temperature can be estimated as~\cite{Abramowicz:2011xu} 
\begin{equation}
T_{\rm disk} \simeq 4\times 10^{3}\alpha^{-1/4}\Big(\frac{M_{\odot}}{M}\Big)^{1/4}\Big(\frac{r}{M}\Big)^{-3/8} \text{eV} \ll m_e, 
\end{equation}
where $\alpha\sim {\cal O}(1)$ is a dimensionless coefficient relating the kinematic viscosity of the fluid with the velocity of turbulent elements.
Therefore, $v_{\rm thermal}\approx 0.06$ or smaller. On the other hand, for a quasibound state around a BH of mass $M$, $\omega\sim \omega_{\rm pl}$ and $k=2\pi/\lambda$, where $\lambda\sim M/(M\omega_{\rm pl})^2$ is the typical length scale of the mode~\cite{Arvanitaki:2010sy,Brito:2014wla}. This gives $\omega/k\sim 0.3$ or larger for the most interesting case $M\omega_{\rm pl}\lesssim 0.5$.
We also stress that in the present work we treat the plasma as spherically symmetric, and static. The first is a simplifying assumption and should be relaxed to accommodate more realistic accretion disk geometries. 
Once spherical symmetry is assumed, the plasma is static to very good approximation: the time scales of interest to this work are much shorter than the time scales of other important astrophysical phenomena. In particular, the BH accretion timescale can be conservatively estimated to be
given by a fraction of the Eddington accretion timescale $\tau_{\rm accr} = M/\dot{M} \sim f_{\rm Edd}^{-1} \, 10^{15}\,{\rm s}$, where $f_{\rm Edd} $ is an accretion efficiency factor. For $f_{\rm Edd} \sim \mathcal{O}(1)$, which is conservative as accretion may be much less efficient, one sees that the time scale of the plasma radial motion is much longer than the time scale of the quasi-trapped long-lived perturbations studied below, $\tau_{\rm accr} \gg \omega_{\rm pl}^{-1}$.

The remaining of this manuscript is organized as follows. In Sec.~\ref{sec:setup}
we define the general setup of our work, the main equations in spherical symmetry (Sec.~\ref{sub:spherical}) and the plasma profiles we considered (Sec.~\ref{sub:plasmaprofiles}).
We show our numerical results for the quasibound states in Sec.~\ref{sec:results} and comment on the implications for spinning BHs in Sec.~\ref{sec:spin}. Finally, we conclude in Sec.~\ref{sec:discussion}.
%
%----------------------------------------------------------------------------------------------------
\section{Setup}
\label{sec:setup}
%----------------------------------------------------------------------------------------------------

\subsection{General equations}
\label{sub:general equation}

The general study of the propagation of EM waves through a cold plasma in curved spacetime was pionereed in~\cite{1981A&A....96..293B}, where the authors derived the system of nonlinear equations governing the plasma and the EM field. In this work we start from those equations and specialize to a background Schwarzschild metric, for which we study the quasibound states for different plasma configurations.

Consider a two-component plasma made of electrons and ions. Let us denote the number density and four- velocity of the electrons as $n$ and $u^{\mu}$, while $J^{\mu}$ stands for the ion current density. The system of differential equations for the plasma quantities reads~\cite{1981A&A....96..293B} 
\begin{align}
\nabla_{\nu} F^{\mu\nu} = e n u^{\mu} + J^{\mu}, \label{eq:Maxwell} \\
u^{\mu} \nabla_{\mu} u^{\nu} = e/m_e F^{\nu}{}_{\mu}u^{\mu} ,\label{eq:momentum}\\
u^{\mu}u_{\mu} = -1, \\
\nabla_{\mu}(nu^{\mu}) = 0. \label{eq:last}
\end{align} 
These are Maxwell's equations together with the momentum and particle conservation equations, in covariant form.

We study the propagation of a perturbation through the plasma by introducing the small perturbations $\tilde{n}, \tilde{u}^{\mu}, \tilde{F}_{\mu\nu}$, e.g. ${F}_{\mu\nu}={F}_{\mu\nu}^{\rm background}+\tilde{F}_{\mu\nu}$ (and likewise for other quantities). Here we neglect second-order perturbations of the plasma and EM field, as well as any perturbation of the background metric $g_{\mu\nu}$ (since the gravitational backreactions of these fields is small). We also neglect perturbations of the ions, since they will be suppressed with respect to those of the electrons by a factor $\propto m_e/ m_{\rm ion} \ll 1$. In the hot plasma case an extra term would appear in the momentum equation \eqref{eq:momentum}, due to the pressure of the fluid.

The presence of a plasma implies the existence of a preferred rest frame. Locally, the plasma defines surfaces of simultaneity for the observer, whose effective metric tensor is 
\begin{equation}
h_{\mu\nu} = g_{\mu\nu} + u_{\mu}u_{\nu}.
\end{equation}
The tensor $h_{\mu\nu}$ projects orthogonally onto the tangent rest plane of the (electron) plasma. 
Then, the kinematic of an electron fluid is described by two matrices: the rate of rotation (the vorticity) $\omega^{\mu\nu} \equiv - \omega^{\nu\mu}$, and the rate of deformation $\theta^{\mu\nu} \equiv \theta^{\nu\mu}$. This follows from the general decomposition~\cite{Ellis:1971pg} 
\begin{equation}
\nabla_{\mu} u_{\nu} =\nabla {}_{(\mu}u_{\nu)} + \nabla {}_{[\mu}u_{\nu]}  ={\omega_{\nu}}_{\mu} + \theta_{\nu\mu} - 
u_{\mu} u^{\alpha}\nabla_{\alpha} u_{\nu}\,,
\end{equation} 
from which we get 
%%%
\begin{eqnarray}
 \omega_{\mu\nu} &=& \frac{1}{2}(v_{\mu\nu}-v_{\nu\mu})\,, \label{vort}\\
 \theta_{\mu\nu} &=& \frac{1}{2}(v_{\mu\nu}+v_{\nu\mu})\,, \label{deform}
\end{eqnarray}
%%%
where we defined the tensor $v^{\mu\nu}=h^{\mu \alpha }h^{\nu \beta }u_{\alpha;\beta}$.
We can also define the plasma frequency as in Eq.~\eqref{plasmafreq}, the electric component $E^{\mu} \equiv 
F^{\mu}{}_{\nu}u^{\nu}$, the magnetic component $B_{\mu\nu} \equiv h_{\mu}{}^{\alpha}h_{\nu}{}^{\beta}F_{\alpha \beta}$, and the Larmor 
tensor ${\omega_{\rm L}}^{\mu\nu} = -\frac{e}{m_e} B^{\mu\nu}$. With these definitions, by differentiating Maxwell's equation  \eqref{eq:Maxwell} and using the momentum equation \eqref{eq:momentum}, Ref.~\cite{1981A&A....96..293B} obtained the perturbed equation for the vector 
potential perturbation $\tilde{A}^{\mu}$ in the Landau gauge, $u_{\mu}\tilde{A}^{\mu} = 0$, containing both the influence of the 
gravitational potential and that of the moving plasma:
\begin{align}\label{eq:final}
\Big(  h^{\alpha\beta}u^{\nu}\nabla_{\nu}(\nabla^{\mu}{}_{\beta} - \delta^{\mu}{}_{\beta}\nabla^{\gamma}{}_{\gamma})+(\omega^{\alpha\beta} + \omega_{\rm L}^{\alpha\beta}+\theta^{\alpha\beta}&   \\
  + \theta h^{\alpha\beta}+ \frac{e}{m_e} E^{\alpha}u^{\beta}) 
\times (\nabla^{\mu}{}_{\beta}- \delta^{\mu}{}_{\beta}\nabla^{\gamma}{}_{\gamma})&\nonumber \\    + \omega_{\rm pl}^2h^{\alpha\mu}u^{\gamma}\nabla_{\gamma} + \omega_{\rm pl}^2(\theta^{\alpha\mu} - \omega^{\alpha\mu}) &\Big)\tilde{A}_{\mu} = 0,\nonumber 
\end{align}
where $\nabla_{\mu\nu} \equiv \nabla_\mu \nabla_\nu$ and $\theta = \theta^{\mu}{}_{\mu}$. 
The above equation is the starting point for a rigorous analysis of the linearized photon dynamics in a cold plasma in curved spacetime.
It is clearly very different from an effective Proca equation which would have the form $\nabla_\alpha\nabla^\alpha \tilde A_\mu=\omega_{\rm pl}^2 \tilde A_\mu$. In particular, note that Eq.~\eqref{eq:final} contains {\it third-order} derivatives. 

In the flat spacetime limit and in the Fourier domain, the spatial part of Eq.~\eqref{eq:final} reads
\begin{equation}
 \bm{k}(\bm{k}\cdot \tilde{\bm{E}})-k^2\tilde{\bm{E}}+\Big(1-\frac{\omega_{\rm pl}^2}{\omega^2}\Big)\omega^2\tilde{\bm{E}}=0\,,
\end{equation}
which is indeed the equation that regulates the propagation of an EM wave in an isotropic plasma with dielectric tensor $\epsilon= 1-\omega_{\rm pl}^2/\omega^2$. Taking $\bm{k}$ along the $z$ direction one can write the standard dispersion tensor $\bm{D}$ and find its determinant, which reads $\text{det} |\bm{D}| = ( \omega^2 \epsilon-k^2)^2 \omega^2 \epsilon = 0$. There are two types of solutions to this dispersion relation: the first one corresponds to the longitudinal mode with $\epsilon = 0$, i.e. $\omega = \omega_{\rm pl}$; while the second one corresponds to the two transverse modes $(\omega^2 \epsilon-k^2) = 0$, i.e. $\omega^2 = k^2 + \omega_{\rm pl}^2$~\cite{Raffelt:1996wa}.

%%%%%%
\subsection{Spherically symmetric spacetimes}
\label{sub:spherical}
%%%%%%%%
We now specialize to the symmetries of the Schwarzschild background. We work in the coordinates $(t,r, \theta,\phi)$, in which the line element reads
\begin{equation}
    ds^2 = - f dt^2 + f^{-1} dr^2 + r^2 d\Omega_2^2
\end{equation}
with $f(r) = 1- 2 M /r$, where $M$ is the BH mass.
In this case both the vorticity and the deformation tensors are zero, as can be easily checked from Eqs.~\eqref{vort} and~\eqref{deform}. 
The four velocity of a static plasma 
is $u^\alpha=(u^0, \vec{0})$, with $ u^0=f^{-1/2}$ satisfying the normalization condition $u_{\mu}u^{\mu} = -1$. From Eq.~\eqref{eq:momentum}, the electric field has then only one nonvanishing radial component $E^\alpha=(0, m_e/e \, \Gamma^r_{00}(u^0)^2,0,0)$, where $\Gamma^\mu_{\alpha\beta}$ are the standard Christoffel's symbols. We assume an unmagnetized plasma $B_{\mu\nu}=0$ (and therefore also ${\omega_{\rm L}}^{\mu\nu}= 0$). 

Moreover, in any spherically symmetric spacetime it is possible to separate the angular part of the fields from the radial one by performing a multipolar expansion. Following Ref.~\cite{Rosa:2011my}, we introduce a basis of four vector spherical harmonics:
\begin{align}
    Z_\mu^{(1)lm}&=[1,0,0,0]Y^{lm}\label{eq:Sphericalh1} ,\\
    Z_\mu^{(2)lm}&=[0,f^{-1},0,0]Y^{lm}, \label{eq:Sphericalh2} \\
    Z_\mu^{(3)lm}&=\frac{r}{\sqrt{l(l+1)}}[0,0,\partial_{\theta},\partial_{\phi}]Y^{lm}, \label{eq:Sphericalh3}\\
    Z_\mu^{(4)lm}&=\frac{r}{\sqrt{l(l+1)}}[0,0,\frac{\partial_\phi}{\sin\theta},-\sin\theta\partial_{\theta}]Y^{lm} \label{eq:Sphericalh4}, 
\end{align}
where $Y^{lm} (\theta, \phi)$ are the standard scalar spherical harmonics. These vector spherical harmonics satisfy the orthogonality condition
\begin{equation}
    \int d\Omega Z^{(i)lm}_{\mu}\hat\eta^{\mu\nu}Z^{(i')l'm'}_{\nu}=\delta^{ii'}\delta^{ll'}\delta^{mm'} ,
\end{equation}
where $d\Omega=\sin\theta d\theta d\phi$ and $\hat\eta^{\mu\nu}={\rm diag}[1,f^2,1/r^2,1/(r^2\sin^2\theta)]$. The perturbation of the vector potential can be decomposed in this basis as
\begin{equation}
    \tilde A_\mu(r,t,\theta,\phi)=\frac{1}{r}\sum_{i=1}^4\sum_{l,m}c_iu_{(i)}^{lm}(t,r)Z_\mu^{(i)lm}(\theta,\phi),
\end{equation}
where $c_1=c_2=1$ and $c_3=c_4=1/\sqrt{l(l+1)}$. Using this decomposition and a frequency-domain representation $u_{(i)}^{lm}(t,r)=u_{(i)}^{lm}(r)e^{-i\omega t}$, the field equations become 
\begin{align}
    u_{(1)}&=0\,, \label{eq:decomposedset1} \\
    \left(r f (l + l^2 + r^2 \omega_{\rm pl}^2) - r^3 \omega^2\right) u_{(2)}
                  - r^2 f^2 u_{(3)}' &=0\,, \label{eq:decomposedset2}\\
                l (1 + l) rf u_{(2)}
                + r^3 ( \omega^2 -f \omega_{\rm pl}^2 ) u_{(3)}  &\nonumber\\- l (1 + l) r^2 f u_{(2)}'+ 
      2 M rf u_{(3)}' + 
       r^3 f^2 u_{(3)}''&=0\,, \label{eq:decomposedset3}\\
\left(r f (l + l^2 + 
         r^2 \omega_{\rm pl}^2) + r^3 \omega^2\right) u_{(4)} \nonumber \\-  2 M r f u_{(4)}' -r^3 f^2  
        u_{(4)}'' &=0\,, \label{eq:decomposedset4}
\end{align}
where $u_{(i)}'=\partial_r u_{(i)}$, we have suppressed the $l$ superscript, and the radial dependence of $\omega_{\rm pl}=\omega_{\rm pl}(r)$. Owing to the spherical symmetry of the background, the equations do not depend on the angular number $m$. Note that, despite the fact that Eq.~\eqref{eq:final} contains third-order derivatives, the final system of equations in the frequency domain is of second differential order.

From Eqs.~\eqref{eq:decomposedset1}--\eqref{eq:decomposedset4} we can immediately notice that the polar (i.e., even-parity) sector, described by the functions $u_{(1)}, u_{(2)}$ and $u_{(3)}$ is completely decoupled from the axial (i.e., odd-parity) sector, described by the function $u_{(4)}$. This resembles the case of a massive vector field discussed in Ref.~\cite{Rosa:2011my} and is in fact a consequence of the spherical symmetry of the background.
We can therefore treat the two sectors independently, as in the following.

\subsubsection{Axial sector}
Interestingly, the axial equation is identical to the axial Proca equation found in~\cite{Rosa:2011my}. 
Thus, the axial sector can be easily reduced to a Schr\"odinger-like equation analogous to the massive vector case~\cite{Rosa:2011my}
\begin{equation}
\label{eq:axial_const}
    \mathcal{D}_2 u_{(4)}(r)=0,
\end{equation}
with $\mathcal{D}_2 \equiv   \frac{d^2}{dr_*^{2}}+ \omega^2- f \Big(\frac{l(l+1)}{r^2}+\omega_{\rm pl}^2\Big)$, and in terms of the tortoise coordinate defined by $\frac{d r_*}{dr} = f^{-1}$. In this case the plasma frequency plays indeed the role of an effective mass for the component $u_{(4)}$.

\subsubsection{Polar sector}
On the other hand, the polar sector is different from the Proca case.

In the monopole case ($l=0$), only the first two spherical harmonics are defined and the field equations reduce to $u_{(1)}=0$ and
%%%
\begin{equation}
    (\omega^2-\omega_{\rm pl}^2 f ) u_{(2)}=0.
\end{equation}
%%%
In general, this equation admits only the trivial solution $u_{(2)}=0$, whereas in the flat spacetime limit ($M\rightarrow 0$) we recover the presence of electrostatic degrees of freedom, $\omega^2=\omega_{\rm pl}^2$, which are typical of photons propagating in cold plasma. 

For $l\neq0$, the polar equations for $u_{(2)}$ and $u_{(3)}$ can be reduced to a single Schr\"odinger-like equation,
\begin{equation}
\label{eq:boundarys}
    \frac{d^2}{dr_*^{2}} \psi-V(r)\psi=0,
\end{equation}
where the complicated form of the effective potential $V(r)$ is given in Appendix~\ref{app:Polar sector}. As expected, $V(r\to2M)\to -\omega^2$ and $V(r\to\infty)\to\omega_{\rm pl}^2(r\to\infty)-\omega^2$. The fact that the polar sector reduces to a single second-order differential equations implies the presence of only one dynamical degree of freedom. Note that the potential $V$ depends on the plasma frequency $\omega_{\rm pl}(r)$ and also on its radial derivatives.

\subsection{The hydrogenic spectrum for Proca modes}

In the next section we shall compute the quasibound states of our problem by solving Eqs.~\eqref{eq:axial_const} and \eqref{eq:boundarys} numerically with suitable boundary conditions. The latter select an infinite set of complex eigenfrequencies: $\omega=\omega_R+i\omega_I$. It will be useful to compare the results with the case of a Proca field with mass $\hbar\omega_{\rm pl}={\rm const}$ around a Schwarzschild BH. In the latter case, the spectrum has a clear physical interpretation in the Newtonian limit, which corresponds to the Compton wavelength of the Proca field, $\sim 1/\omega_{\rm pl}$, being much larger than the horizon size. This requires $M\omega_{\rm pl}\ll1$. To leading order in this limit the spectrum of quasibound states has a hydrogenic form~\cite{Galtsov:1984ixy,Rosa:2011my,Pani:2012vp,Pani:2012bp,Baryakhtar:2017ngi} 
%%%
\begin{eqnarray}
 \omega_R &\sim&\omega_{\rm pl}\left(1-\frac{(M\omega_{\rm pl})^2}{2(l+S+1+n)^2}\right)\,,  \label{hydrogenic}\\
 \omega_I &\sim& -\frac{1}{2}C^{(1)}_{lS} (M\omega_{\rm pl})^{4l+2S+5} \omega_R\,,\label{wIslope}
\end{eqnarray}
%%%
where $l$ is the total angular momentum\footnote{This can be seen by applying the angular-momentum operator to the state~\cite{Baryakhtar:2017ngi}. Note that $l$ was erroneously identified with the orbital angular momentum in Ref.~\cite{Rosa:2011my}. See Ref.~\cite{Brito:2015oca} for a discussion of different spherical bases used in the literature.} 
of the state with spin projections $S=-1$, 0, 1 (with $S=0$ for axial modes and $S=\pm1$ for the two polarizations of polar modes), $n$ is the overtone number ($n=0$ for the longest-lived, fundamental mode), and $C^{(1)}_{lS}$ are constants (given, e.g., in Ref.~\cite{Brito:2015oca}). The most unstable mode is the polar dipole with $S=-1$, $l=1$, which has $C^{(1)}_{1-1}=16$~\cite{Baryakhtar:2017ngi}.
The dominant slope of the imaginary part of the mode is $\omega_I \propto (M\omega_{\rm pl})^{10}$ for the axial dipole, and $\omega_I \propto (M\omega_{\rm pl})^8$ and $\omega_I \propto (M\omega_{\rm pl})^{12}$ for the two polar dipole modes. 
The above analytical approximation in the Newtonian limit is in excellent agreement with the exact numerical 
results~\cite{Detweiler:1980uk,Dolan:2007mj,Baryakhtar:2017ngi,Cardoso:2018tly,Dolan:2018dqv,Baumann:2019eav,Brito:2013wya,Brito:2020lup}.

In the spinning case, the imaginary part acquires a factor $\omega_I\propto(\omega_R-m\Omega)$, which depends on the BH angular velocity $\Omega$. Therefore, in the superradiant regime, $\omega_R<m\Omega$, the modes with the smallest slope in the static case (namely, the polar dipole with $S=-1$) become the ones with the shortest instability timescale, $\tau=1/\omega_I$~\cite{Brito:2015oca} (see Sec.~\ref{sec:spin} for more details).

In the next section we shall compare our numerical results for the full plasma-photon system with the above hydrogenic behavior of a Proca field.

\subsection{Plasma profiles}
\label{sub:plasmaprofiles}

We consider two different plasma profiles: a homogeneous density profile and a Bondi-like accretion disk model.
We will examine first the case of homogeneous density, and therefore homogeneous plasma frequency, $\omega_{\rm pl}={\rm const}$. This approximation is obviously not realistic, especially close to the BH. However, it will serve as a warm-up to elucidate the structure of the equations and identify the correct limit asymptotically far from the BH. 

Then, we will consider a radial dependence $\omega_{\rm pl}=\omega_{\rm pl}(r)$, as expected in the surroundings of a spherically symmetric BH. In particular, we will consider the Bondi-like accretion disk model, which describes the accretion dynamics of a non-self-interacting gas around a spherically symmetric compact object~\cite{1952MNRAS.112..195B}. This model predicts a power-law density profile for the gas and consequently a plasma frequency of the type 
\begin{equation}
    \omega_{\rm pl}^2(r) = \omega_{\rm B}^2\Big(\frac{2M}{r}\Big)^{\lambda} + \omega_{\rm \infty}^2, \label{Bondi}
\end{equation}
where $\sqrt{\omega_{\rm B}^2+\omega_\infty^2}$ is the plasma frequency at the horizon (since in the relevant regime $\omega_\infty\ll\omega_{\rm B}$, with a little abuse of notation we shall refer to $\omega_{\rm B}$ as the horizon plasma density). The slope $\lambda$ depends on the adiabatic index of the gas (e.g., $\lambda = 3/2$ for monoatomic species). The constant term $\omega_{\infty}$ is the asymptotic plasma frequency at infinity, i.e the interstellar medium plasma frequency far away from the central BH.

%----------------------------------------------------------------------------------------------------
\section{Plasma-driven quasibound EM modes of a Schwarzschild BH}
\label{sec:results}
%----------------------------------------------------------------------------------------------------

\subsection{Numerical method}
\label{sec:numerics}

We compute the characteristic frequencies of our system using a
direct integration shooting method~\cite{Ferrari:2007rc, Pani:2012bp, Rosa:2011my}. The main idea is to integrate the system from the horizon outwards to infinity, imposing suitable boundary conditions. Close to the horizon, the solution must be a purely ingoing wave,
as the horizon behaves as a one-way membrane,
\begin{equation}
\label{eq:boundaryh}
    u_{(i)}\sim e^{-i\omega r_*}\sum_n b_{(i)\, n} (r-2M)^n,
\end{equation}
where the coefficients $b_{(i)\, n}$ can be computed in terms of the arbitrary coefficient $b_{(i)\, 0}$ by expanding the relevant equations near the horizon.
At infinity, the solution at the leading order can be written in its most generic form as
\begin{equation}
u_{(i)}\sim B_{(i)}e^{-k_{\infty}r_* }+C_{(i)}e^{+k_{\infty}r_*},
\end{equation}
where $k_{\infty}=\sqrt{\omega_{\rm pl}^2(r\to\infty)-\omega^2}$.
We are interested in strongly localized, quasibound state solutions for $\omega<\omega_{\rm pl}$, and we will therefore impose the condition $C_{(i)}=0$, implying exponentially damped solutions at infinity, and solve the associated eigenvalue problem. This method does not rely on a specific shape of the effective potential and is thus extremely flexible. Moreover, the direct integration method works particularly well for the computation of quasibound
states, while it is expected to be less precise for the computation of normal modes (corresponding to the boundary condition $B_{(i)} = 0$), especially for the highly damped modes.

Since the axial and polar sectors are decoupled and each one is described by a single second-order differential equation, for a given value of $l$ we expect to find two families of modes (axial and polar), each one defined by an overtone number $n$. In the following we shall mainly focus on the fundamental ($n=0$) modes, although tracking a fixed overtone number is difficult in the small $M\omega_{\rm pl}$ regime.
As we shall see, in the same regime the imaginary part of the mode can become extremely small and sensitive to small numerical errors. The latter can be reduced by increasing the numerical accuracy and the truncation order of the series expansions at the horizon and at infinity.

This procedure is also suitable to find unstable modes, i.e. those with $\omega_I>0$ and whose time dependence is exponentially growing as $e^{\omega_I t}$. We have searched for such modes in different configurations and found none, confirming the reasonable expectation that the system is stable in the static case.

\subsection{Constant-density plasma}

\begin{figure}[t]
\centering
\includegraphics[width=0.5\textwidth]{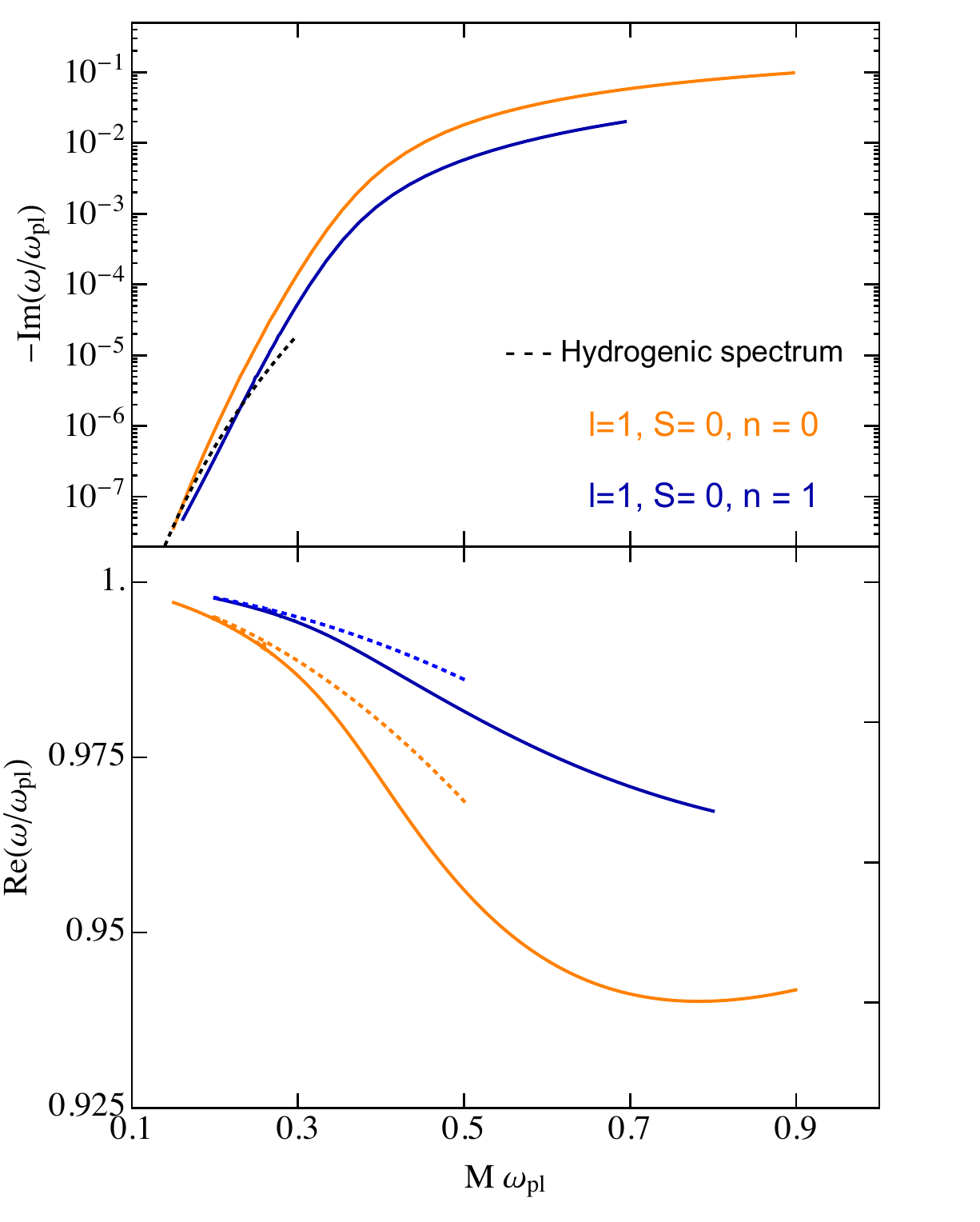}
\caption{Imaginary (upper panel) and real (lower panel) part of the plasma-driven quasibound EM mode of a Schwarzschild BH in the axial sector for the case $\omega_{\rm pl} = {\rm const}$. The imaginary part of the frequency is always negative, while the real part is always smaller than the plasma frequency. We show the fundamental mode ($n=0$, orange) and the first overtone ($n=1$, blue) for $l=1$. The dashed curves correspond to the hydrogenic spectrum, which is recovered as expected in the limit $M \omega_{\rm pl} \ll 1$. Note that the axial sector is equivalent to that of a Proca field~\cite{Rosa:2011my}.} 
\label{fig:axial_spectra_const}
\end{figure}

Figure~\ref{fig:axial_spectra_const} shows the real part and the imaginary part of the axial eigenvalues $\omega=\omega_R+i\omega_I$, normalized to the plasma frequency. We consider $l=1$ and two different overtone numbers, respectively $n=0$ (orange) and $n=1$ (blue). 
As expected, the imaginary part of the frequency is always negative and represents a mode which is exponentially decaying in time, while the real part is always smaller than $\omega_{\rm pl}$, a necessary condition to obtain solutions that are confined in the vicinity of the BH.
Since the axial sector is the same as in the Proca case, this plot is equivalent to what found in Ref.~\cite{Rosa:2011my} for the axial quasibound states of a Proca field around a Schwarzschild BH.
The dashed curves shown in Fig.~\ref{fig:axial_spectra_const} correspond to the hydrogenic spectrum expected in the $M\omega_{\rm pl }\ll1$ limit.

\begin{figure}[t]
\centering
\includegraphics[width=0.5\textwidth]{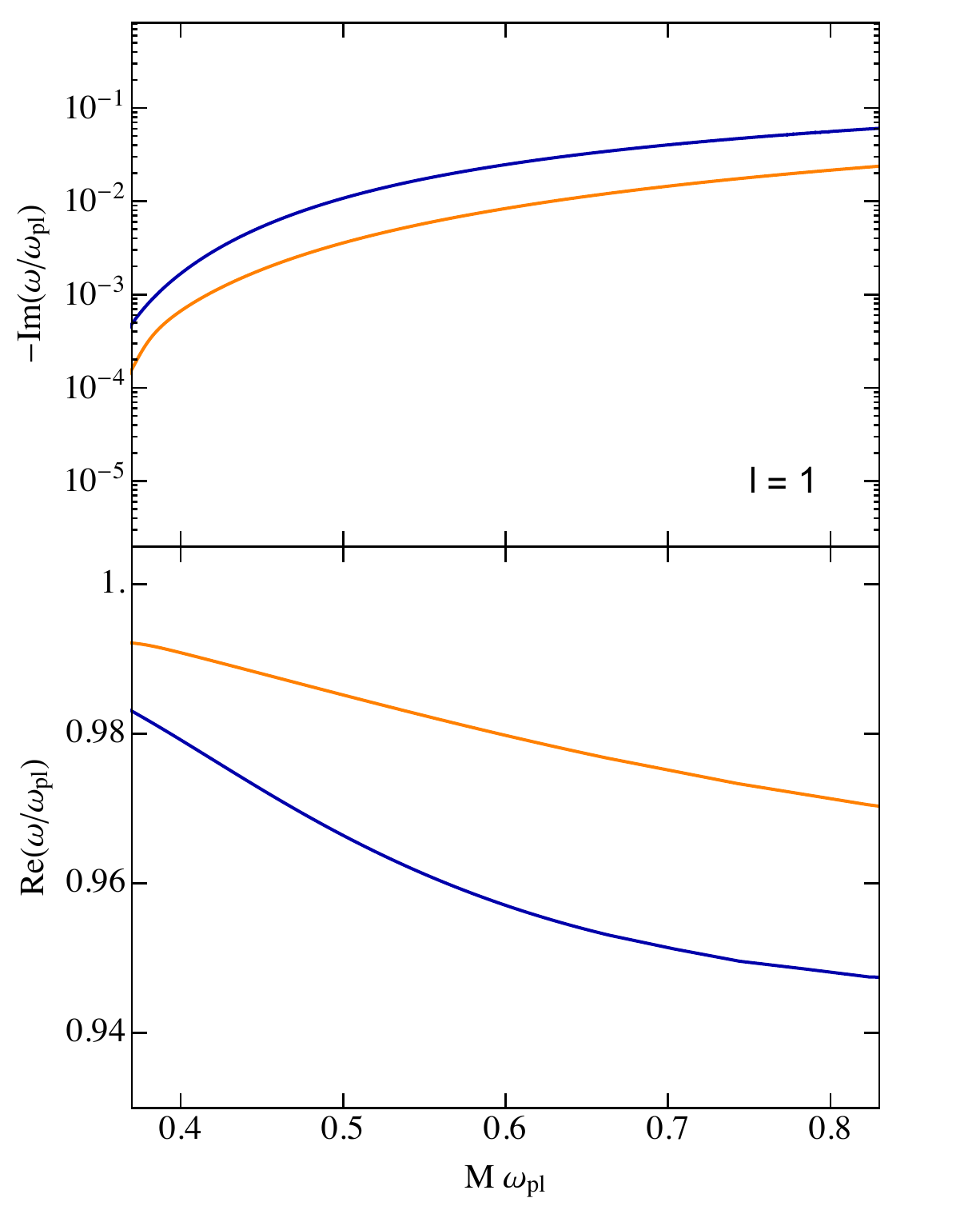}
\caption{Same as in Fig.~\ref{fig:axial_spectra_const} but for the polar sector. In this case the modes do not reduce to the hydrogenic Proca spectrum in the $M\omega_{\rm pl}\ll1$ limit.} 
\label{fig:polar_spectra_const}
\end{figure}

While in the axial sector the potential is simple and equivalent to that of a Proca equation, in the polar sector the situation is different. Indeed, not only are the equations more involved than in the Proca case, but the polar sector only contains a single propagating mode, in contrast with the two polar modes of a Proca field.
Thus, as expected, the structure of the plasma-driven polar quasibound states is different and cannot be mapped into a hydrogenic spectrum as in the axial case. This is shown in Fig.~\ref{fig:polar_spectra_const}.
%%%
%%%%

\begin{figure}[t]
\centering
\includegraphics[width=0.45\textwidth]{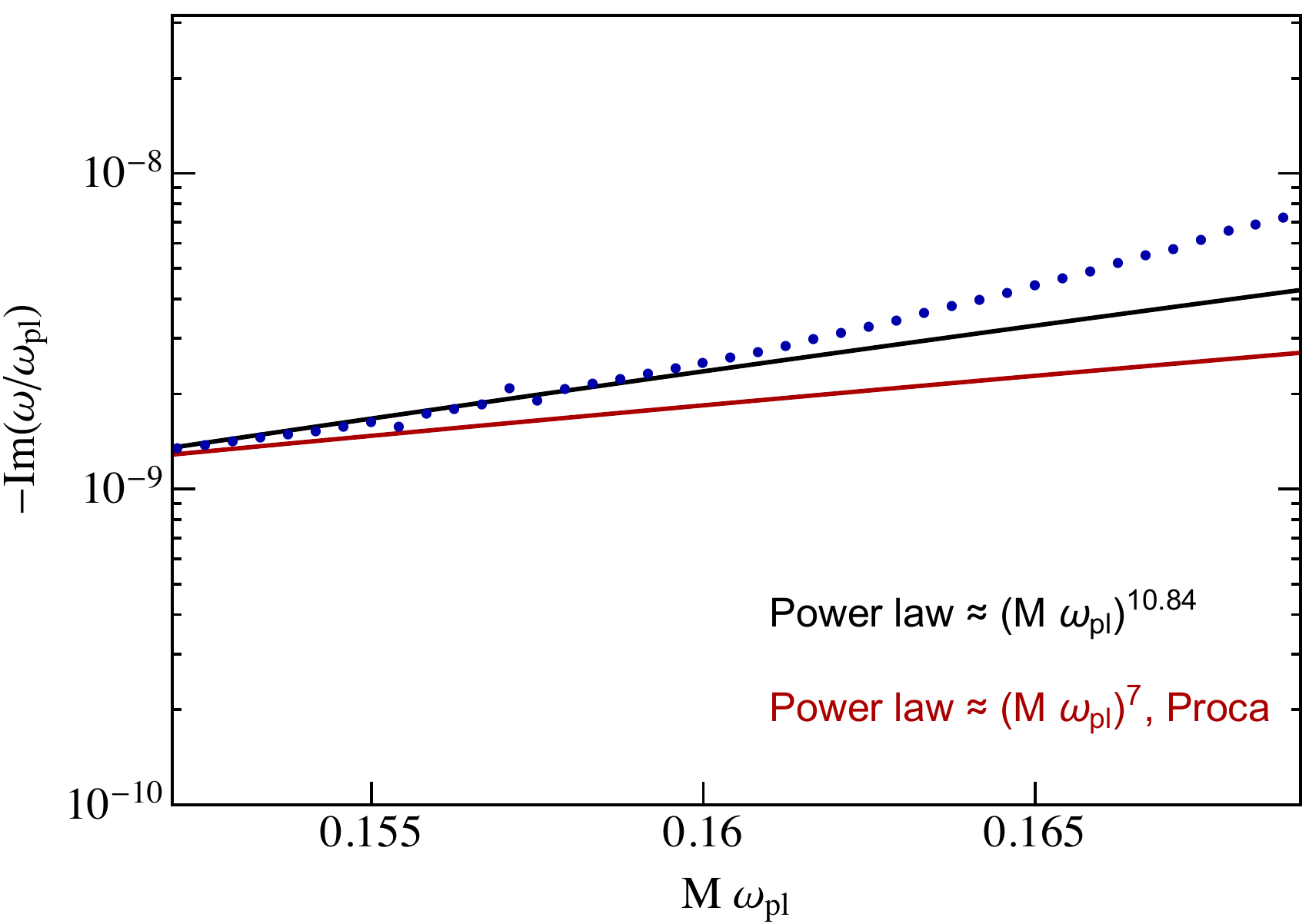}\hfill
\caption{Imaginary part of the plasma-driven EM quasibound mode in the polar sector in $M \omega_{\rm pl} \ll 1$ regime. The blue dots indicate the numerical data; the black solid line is the best-fit power law, while the red line denotes the hydrogenic scaling in the Proca case.} 
\label{fig:polar_const_small}
\end{figure}

In this case the numerical computation of the eigenfrequencies becomes
increasingly more challenging for values of the coupling $M\omega_{\rm pl} \lesssim 0.4$ and we therefore do not show the lower part of the spectrum in Fig.~\ref{fig:polar_spectra_const}. However, in order to investigate the small-mass coupling regime, $M\omega_{\rm pl} \ll 1$, we separately studied smaller intervals around $M\omega_{\rm pl} \simeq 0.15$, as shown in Fig.~\ref{fig:polar_const_small}.
Although the numerical results are noisy in this limit, we find that the scaling of $\omega_I$ is significantly different from the hydrogenic behavior predicted in the Proca case for $S=-1$. Indeed, our best fit yields $\omega_I/\omega_{\rm pl}\sim (M\omega_{\rm pl})^{10.84}$, to be compared with $\omega_I/\omega_{\rm pl}\sim (M\omega_{\rm pl})^{7}$ for the $S=-1$ Proca mode. In fact, we find a scaling that is quite close to the (subleading) polar mode with polarization $S = 1$ for the massive vector case~\cite{Rosa:2011my}.

%%%%%

In some sense, the plasma-driven EM quasibound states are phenomenologically closer to a massive scalar field than to a Proca field. In both plasma-driven EM fields and massive scalar fields, the minimum scaling of the imaginary part (which sets the shortest, and therefore most interesting, mode lifetime) is $\omega_I/\omega_{\rm pl}\sim (M\omega_{\rm p})^{9}$. This scaling is set by the {\rm axial} dipole mode in the plasma-photon system. Thus, the lifetime of the fastest quasibound states in the plasma-driven case is parametrically longer than for a Proca field -- for which the shortest lifetime is set by the polar modes -- by a factor $(M\omega_{\rm pl})^{-2}\gg1$.

\subsection{Plasma profile from Bondi accretion}

We now analyze the quasibound states for a more realistic plasma density profile, as predicted by Bondi accretion. 
In this case the axial sector is also described by Eq.~\eqref{eq:axial_const}, except for a radial dependence in the plasma frequency 
$\omega_{\rm pl} \rightarrow \omega_{\rm pl}(r) = \sqrt{\omega_{\rm B}^2(2M/r)^\lambda + \omega_{\infty}^2}$.
The polar sector is again more involved and its potential depends nontrivially
on the radial derivatives of the plasma frequency. The equations for the polar case are presented in Appendix~\ref{app:Polar sector}.

In Fig.~\ref{fig:axial_Bondi} we show the results for the imaginary and real part of the frequency in the axial sector as a function of $\omega_\infty$, the quantity that would asymptotically corresponds to a mass term. In all cases we fix $\lambda = 3/2$ for the Bondi model (corresponding to monoatomic gas) and we show the results for different values of the horizon plasma frequency $\omega_{\rm B}$. We notice that, as in the homogeneous plasma case, the real (imaginary) part of the frequency decreases (increases) monotonically with $M\omega_{\infty}$. The effect of $\omega_{\rm B}$ is the opposite: larger values of the density in the vicinity of the BHs lead to a smaller imaginary part and a larger real part. Indeed, for larger values of $\omega_{\rm B}$ it is more difficult to find minima in the effective potentials, i.e. it is more difficult to support quasibound states. In fact, we were not able to find quasibound states solutions for $\omega_{\rm B} M  \gtrsim  1.7 $, in good agreement with the numerical study in Ref.~\cite{Dima:2020rzg} for a scalar toy model. Furthermore, this behavior is consistent with the $\omega_B\rightarrow 0$ limit, as in this case the spectrum shown in Fig. \ref{fig:axial_spectra_const}, with smaller values of the real part of the eigenfrequencies, must be recovered.

\begin{figure}[t]
\centering
\includegraphics[width=0.55\textwidth]{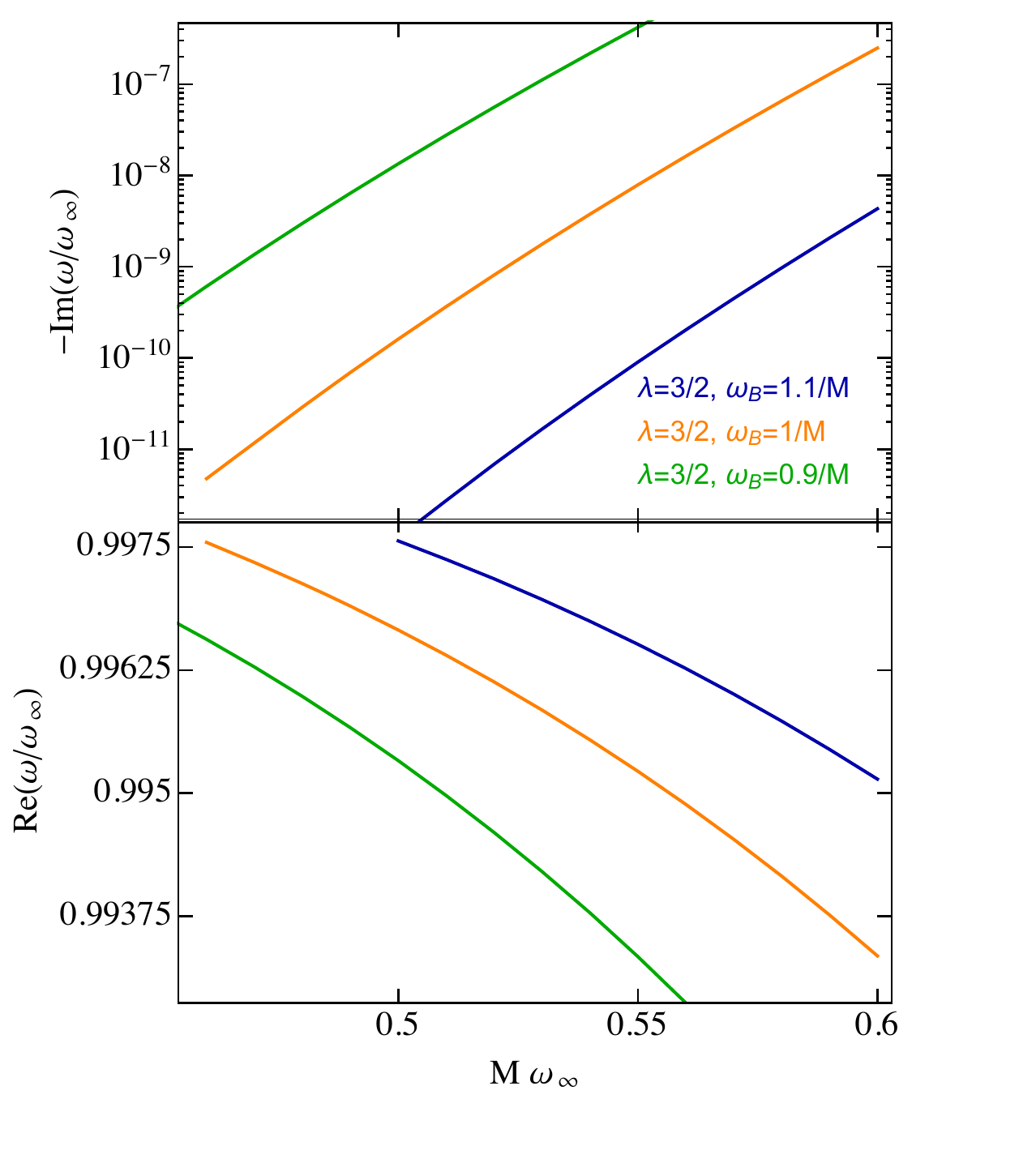}\\
\caption{Imaginary (upper panel) and real (lower panel) part of the quasibound state frequency in the axial sector for a photon in a Bondi-accretion plasma model, see Eq.~\eqref{Bondi}. The imaginary part is always negative, while the real part is always smaller than the plasma frequency, as expected. The imaginary part of the frequency decreases (the modes become short-lived) as the horizon plasma $\omega_{\rm B}$ density increases. }
\label{fig:axial_Bondi}
\end{figure}

Note that, in general, the imaginary part of the quasibound states for a Bondi plasma profile is smaller than in the homogeneous-plasma case. This makes the numerical computation harder, especially for polar modes for which the imaginary part is even smaller than in the axial case. Although not shown, we were also able to compute the plasma-driven polar mode in the case of a Bondi profile, at least when $M\omega_\infty={\cal O}(1)$. As in the homogeneous case, the axial modes are found to be the shortest lived for small values of the plasma frequency.

%----------------------------------------------------------------------------------------------------
\section{The role of the BH spin}
\label{sec:spin}
%----------------------------------------------------------------------------------------------------
In this work we focused on finding the plasma-driven EM quasibound states around a nonspinning (Schwarzschild) BH. A natural and paramount extension is to consider the spinning case, namely solving the equations on a Kerr background metric.

This case is significantly more involved, but also more interesting in view of the fact that the quasibound states can become (superradiantly) unstable in the spinning case~\cite{Brito:2015oca}. A detailed analysis of the photon-plasma interactions around a Kerr BH is left for future work. Here we anticipate the main differences relative to the static case considered here:

\begin{itemize}
    \item {\it Vorticity.} In a Kerr BH the vorticity tensor does not vanish, i.e. (from Eq.~\eqref{vort}) $\omega_{ab}\sim {\cal O}(\chi)$, where $\chi=J/M^2$ is the dimensionless angular momentum of a Kerr BH. On the other hand, the rate of expansion is still zero, $\theta_{ab}=0$. Therefore, the linearized field equations~\eqref{eq:final} contain extra terms which vanish in the static case;
    %%%
    \item {\it Plasma profile.} A spinning BH drags plasma in its vicinity. In the stationary and axisymmetric case, the plasma density can generically depend both on the radius and on the angle $\theta$, i.e. $\omega_{\rm pl}= \omega_{\rm pl}(r,\theta)$;
    %%%
    \item {\it Mode coupling.} In the spinning case, axial and polar modes with different harmonic index $l$ are coupled to each other~\cite{Kojima:1992ie,Pani:2012bp,Pani:2012vp,Pani:2013pma}. Thus, even if the axial sector is the same as in the Proca case when the BH is static, in the spinning case it would get corrections arising from the different polar sector. For a Kerr BH we therefore expect that the {\it entire} quasibound spectrum will differ from the Proca one.
\end{itemize}

Overall, for the above reasons do not expect the linearized equations to be separable, at variance with the Proca case~\cite{Frolov:2018ezx} (unless possibly for specific profiles, see e.g.~Refs.~\cite{Cardoso:2013opa,Cardoso:2013fwa}). The linear dynamics can still be computed perturbatively in the slow-rotation approximation~\cite{Pani:2012bp,Pani:2012vp,Pani:2013pma}, or fully numerically (see, e.g., Ref.~\cite{Cardoso:2018tly} for the Proca case), or using the methods adopted in Refs.~\cite{Dolan:2012yt,Baumann:2019eav,Dima:2020rzg}. 

It is reasonable to expect that, for a Kerr BH, the quasibound spectrum is modified by corrections proportional to the superradiant~\cite{Brito:2015oca} factor $\omega-m\Omega$, where $m$ is the azimuthal number of the perturbation and $\Omega$ is the BH angular velocity. In particular, the imaginary part of the quasibound state might change sign when $\omega<m\Omega$, turning the mode from stable to unstable. This correction is evident in the hydrogenic spectrum of massive bosonic perturbations of a Kerr BH~\cite{Detweiler:1980uk,Pani:2012bp,Pani:2012vp,Baryakhtar:2017ngi,Cardoso:2018tly,Brito:2020lup} in the (Newtonian) small-mass limit.
Nonetheless, for the reasons listed above one cannot exclude that the quasibound spectrum of a Kerr BH can be {\it qualitatively} different from the Proca case. A detailed analysis is required and will appear elsewhere.

It is important to stress that the plasma-driven superradiant instability can be strongly quenched by nonlinear and relativistic effects~\cite{Cardoso:2020nst,Blas:2020nbs}. Thus, although we expect that the {\rm linear} spectrum of plasma-driven EM modes around a Kerr BH is significantly different than in the Proca case, the nonlinear and relativistic quenching might be efficient also for this more realistic configuration. 

For the Schwarzschild background considered in this work, expanding Eqs.~\eqref{eq:Maxwell}--\eqref{eq:last} to second order in the perturbations, we find that the linearized analysis is valid provided
\begin{equation}
    |\partial_i E^i| \ll e n_e \,.
\end{equation}
This equation implies $ e\omega E/v\ll \omega_{\rm pl}^2 m_e$, where $E$ is the amplitude of the electric field, and $v=\omega/k$ is the velocity of the EM wave in the plasma. For $v\sim1$ and $\omega\sim\omega_{\rm pl}$ we recover the condition $e |E| \ll m_e \omega_{\rm pl}$, which implies that the plasma backreaction on the dispersion relation is negligible. For a stable quasibound state that decays in time, this condition is always satisfied if it holds initially. However, for a superradiantly growing state the condition is eventually violated before the EM field can extract enough energy from the BH~\cite{Cardoso:2020nst}.

%----------------------------------------------------------------------------------------------------
\section{Conclusion and extensions}
\label{sec:discussion}
%----------------------------------------------------------------------------------------------------
We initiated a detailed investigation of the equations governing the linearized dynamics of an EM field propagating in a plasma in curved spacetime. Our main goal was to explore the differences with the case of Proca theory which has been so far used as a proxy for the plasma-photon interaction around a BH. We focused on the case of cold plasma around a background Schwarzschild BH and computed the quasibound spectrum of EM modes.
We showed that, while the axial sector is identical to the Proca one, polar perturbations propagate a single degree of freedom (at variance with the two polar modes of a Proca field) which is described by a new field equation. This drastically affects the polar quasibound-mode spectrum and the scaling of these modes in the small-frequency limit.

We also studied the stability of the system, by looking for unstable modes numerically. We explored a large region of the parameter space and found no unstable modes, as expected in the static case.

As discussed in the previous section a natural and urgent extension is to consider the case of plasma-driven EM modes of a spinning (Kerr) BH, which is likely to be drastically different from the Proca case. We expect quantitatively (and perhaps qualitatively) different results for what concerns the superradiant instability. In this context, our analysis suggests that the linearized analyses performed in the past (e.g.,~\cite{Pani:2013hpa,Conlon:2017hhi,Dima:2020rzg}) in the context of plasma-driven superradiant instabilities around spinning BHs should be revisited.

Although a detailed analysis in the spinning case is required, our results already suggest that the dominant unstable modes could be more akin to the scalar massive modes of a Kerr BH rather than to the vector massive modes. If confirmed, this would imply that the toy model adopted in Ref.~\cite{Dima:2020rzg} could be more accurate than what one would in principle expect.

Another interesting extension of our work is to study nonlinear plasma-photon effects in a coherent and covariant framework. It was recently argued that nonlinear and relativistic effects can strongly quench the superradiant instability~\cite{Cardoso:2020nst,Blas:2020nbs}. The framework adopted here could be used to study this interesting problem quantitatively. For example, one could expand the exact field equations to second order in the perturbations and study the backreaction of the plasma onto the EM field.

We also neglected magnetic fields and focused on the case of cold plasma. Considering a magnetized/hot plasma is another natural extension.

Finally, we assumed that the plasma was static, but this is not the case for an accreting BH. While this approximation is valid for modes that are short-lived with respect to the typically accretion timescale of the BH, this might not always be the case, especially for problems (e.g. the superradiant instability) which could be characterized by extremely long timescales. Thus, another interesting extension would be to solve the linearized photon-plasma dynamics in the time domain, also taking into account the dynamics of the plasma density, or perform a multiscale adiabatic approximation~\cite{Brito:2014wla}.

We hope to report on these interesting problems in the near future.

%----------------------------------------------------------------------------------------------------
\begin{acknowledgments}
%----------------------------------------------------------------------------------------------------
We thank Richard Brito and Vitor Cardoso for useful conversations and comments.
P.P. acknowledges financial support provided under the European Union's H2020 ERC, Starting 
Grant agreement no.~DarkGRA--757480. We also acknowledge support under the MIUR PRIN and FARE programmes (GW-NEXT, CUP:~B84I20000100001), and from the Amaldi Research Center funded by the MIUR program ``Dipartimento di Eccellenza'' (CUP: 
B81I18001170001). A.C. acknowledges support from the the Israel Science Foundation (Grant No. 1302/19), the US-Israeli BSF (grant 2018236) and the German Israeli GIF (grant I-2524-303.7). A.C. acknowledges hospitality from the MPP of Munich. 
L.S.~acknowledges that research at Perimeter Institute is supported in part by the Government of Canada through the Department of Innovation, Science and Economic Development Canada, and by the Province of Ontario through the Ministry of Colleges and Universities.
\end{acknowledgments}

\appendix

\section{Polar sector}\label{app:Polar sector}

Here we derive the full potential for the polar sector. We first consider a homogeneous plasma ($\omega_{\rm pl}={\rm const}$) and then a generic radially dependent density profile, i.e. $\omega_{\rm pl}=\omega_{\rm pl}(r)$.

\subsection{Homogeneous plasma case}

From Eq.~\eqref{eq:decomposedset2}, it is possible to find an expression for $u_{(2)}$ in terms of $u_{(3)}'$. This expression can be inserted in Eq.~\eqref{eq:decomposedset3} to get a single second order equation for the decoupled variable $u_{(3)}$:
\begin{widetext}
\begin{align}
\label{eq:PolarConst}
u_{(3)}''-u'_{(3)} \frac{2 f^2 l (l+1) \omega ^2+(f-1) f^2 r^2 \omega _{\text{pl}}^4-f \omega _{\text{pl}}^2 \left(f (f+1) l (l+1)+2 (f-1) r^2 \omega ^2\right)+(f-1) r^2 \omega ^4}{f r \left(f \omega _{\text{pl}}^2-\omega ^2\right) (f \left(l^2+l+r^2 \omega _{\text{pl}}^2\right)-r^2 \omega ^2)}& \nonumber\\
+ u_{(3)} \left(\omega ^2-  \frac{f \left(l^2+l+r^2 \omega _{\text{pl}}^2\right)}{r^2} \right)f^{-2}&=0.
\end{align}
\end{widetext}
Equation ~\eqref{eq:PolarConst} is still not in the form of a Schroedinger-like equation. We therefore perform the following substitution: $u_{(3)}=G(r)\psi(r)$, where $G$ is an unknown function such that with this transformation equation Eq.~\eqref{eq:PolarConst} assumes the form $d^2\psi/dr_*^2-V(r)\psi=0$. 
We find $G(r) = r^{-1} \sqrt{f r (l^2+l+r^2 \omega_{\rm pl}^2)-r^3 \omega ^2}  \sqrt{r (f \omega_{\rm pl}^2-\omega ^2)}^{-1}$
and we finally arrive at a Schrodinger-like equation in terms of the tortoise coordinate where the potential has the following cumbersome form 
\begin{widetext}
\begin{align}
\label{eq_cfunction}
    & V(r) = \frac{d(r)}{4 \left(\omega ^2-f \omega _{\text{pl}}{}^2\right) \left(f l (l+1) r+f r^3 \omega _{\text{pl}}{}^2-r^3 \omega ^2\right)}\,,
\end{align}
with
\begin{multline}
\label{eq_cfunctiond}
     d(r)  = 
    -\omega ^4 \Big[f^4 \left(3 l^2 (l+1)^2+72 l (l+1) r^2 \omega _{\text{pl}}^2\right)+2 f^3 \Big(-18 l (l+1) \left(l^2+l+1\right) r^2 \omega _{\text{pl}}^2  \\
    -l^2 (l+1)^2 (2 l (l+1)+1)-36 l (l+1) r^4 \omega _{\text{pl}}^4-20 r^6 \omega _{\text{pl}}^6\Big)-f^2 l^2 (l+1)^2\Big]  \\
    -\omega ^8 \left(-12 f l (l+1) r^4-20 f r^6 \omega _{\text{pl}}^2\right)-4 r^6 \omega ^{10} -\omega ^2 \Big[2 f^5 \omega _{\text{pl}}^2 \left(-3 l^2 (l+1)^2-29 l (l+1) r^2 \omega _{\text{pl}}^2\right)  \\
    +f^4 \left(4 l^2 (l+1)^2 (2 l (l+1)+1) \omega _{\text{pl}}^2+48 l (l+1) r^4 \omega _{\text{pl}}^6+4 l (l+1) (9 l (l+1)+5) r^2 \omega _{\text{pl}}^4+20 r^6 \omega _{\text{pl}}^8\right)  \\
    +2 f^3 \left(l^2 (l+1)^2 \omega _{\text{pl}}^2+l (l+1) r^2 \omega _{\text{pl}}^4\right)\Big]-4 f^6 l (l+1) \omega _{\text{pl}}^4 \left(l^2+l+4 r^2 \omega _{\text{pl}}^2\right) \\
    -2 f^5 \omega _{\text{pl}}^2 \left(-2 l^2 (l+1)^2 \left(l^2+l+1\right) \omega _{\text{pl}}^2-6 l (l+1) r^4 \omega _{\text{pl}}^6-2 l (l+1) (3 l (l+1)+1) r^2 \omega _{\text{pl}}^4-2 r^6 \omega _{\text{pl}}^8\right)  \\
    -\omega ^6 \left(f^2 \left(48 l (l+1) r^4 \omega _{\text{pl}}^2+4 l (l+1) (3 l (l+1)+5) r^2+40 r^6 \omega _{\text{pl}}^4\right)-30 f^3 l (l+1) r^2-2 f l (l+1) r^2\right)\,.
    %\bigg(&
     %\left(l(l+1) rf+r^2 (\omega_{\rm pl}^2 rf-r \omega^2)\right) \Big(r (2 M \omega_{\rm pl}^2+r (\omega-\omega_{\rm pl}) (\omega+\omega_{\rm pl}))^2 \nonumber\\
     %& \ \times \left( l(l+1) rf
    %+r^2 (\omega_{\rm pl}^2 rf-r \omega^2)\right)^3 - 2 l (l+1) rf (r \omega^2 (3 M-r)+\omega_{\rm pl}^2 r^2f^2) \nonumber\\
    %&\ \times (r \omega^2 rf (2 M r^2 \omega_{\rm pl}^2-l (l+1) rf) 
    %+\omega_{\rm pl}^2 r^2f^2 (M r^2 \omega_{\rm pl}^2-l (l+1) (M-r))+M r^4 \omega^4)\Big)\bigg) / (rf (l^2+l+r^2 \omega_{\rm pl}^2)+r^3 \omega^2)  \nonumber\\
    %&+ l (l+1) r^2f^2 \Big(l (l+1) r^2 \omega^4 (15 M^2-12 M r+2 r^2) + \omega_{\rm pl}^4 rf (r^3 \omega^2 (-44 M^2+42 M r-9 r^2)
    %+2 l (l+1) r^3f^3) \nonumber\\ &
    %+ r \omega^2 \omega_{\rm pl}^2 (4 l (l+1) rf (5 M^2-5 M r+r^2)+r^3 \omega^2 (56 M^2-48 M r+9 r^2)) -3 r^5 \omega^6 (r-4 M)+3 r^2 \omega_{\rm pl}^6 r^4f^4\Big).
\end{multline}
\end{widetext}
Note that $V\to-\omega^2$ as $r\to2M$ and $V\to \omega_{\rm pl}^2-\omega^2$ as $r\to\infty$.

\subsection{Nonhomogeneous plasma case}

The case of a generic radially dependent plasma density proceeds in the same fashion as the homogeneous case. We find the same expression for the function $G$ with $\omega_{\rm pl}$ replaced by $\omega_{\rm pl}(r)$. 
As in the previous case we then arrive to a Schrodinger-like equation with an effective potential

\begin{widetext}
\begin{align}
\label{eq_cfunction2}
    & V(r) = \frac{d(r)}{4 \left(\omega ^2-f \omega _{\text{pl}}(r){}^2\right) \left(f l (l+1) r+f r^3 \omega _{\text{pl}}(r){}^2-r^3 \omega ^2\right)}\,,
\end{align}
where 
\begin{multline}
\label{eq:cBondi}
 d(r) = \omega ^4 \Big[36 f^3 l (l+1) r^2 \left(-2 f+l^2+l+1\right) \omega _{\text{pl}}(r){}^2+f^2 l^2 (l+1)^2 (f (-3 f+4 l (l+1)+2)+1)\\
 -4 f^3 l (l+1) r \omega _{\text{pl}}(r) \left((9 f-5) r^2 \omega _{\text{pl}}'(r)-f r^3 \omega _{\text{pl}}''(r)\right)\\
 +4 f^4 l (l+1) r^4 \omega _{\text{pl}}'(r){}^2+72 f^3 l (l+1) r^4 \omega _{\text{pl}}(r){}^4+40 f^3 r^6 \omega _{\text{pl}}(r){}^6\Big]+\\
 \omega ^2 \Big[2 f \omega _{\text{pl}}(r){}^2 \left(f^2 l^2 (l+1)^2 (f (3 f-4 l (l+1)-2)-1)+4 f^4 l (l+1) r^4 \omega _{\text{pl}}'(r){}^2\right)\\-4 f^5 l^2 (l+1)^2 r^2 \omega _{\text{pl}}'(r){}^2
 +4 f^4 l (l+1) r \omega _{\text{pl}}(r){}^3 \left(2 (7 f-3) r^2 \omega _{\text{pl}}'(r)-2 f r^3 \omega _{\text{pl}}''(r)\right)\\-48 f^4 l (l+1) r^4 \omega _{\text{pl}}(r){}^6
 -2 f^3 l (l+1) r^2 (f (-29 f+18 l (l+1)+10)+1) \omega _{\text{pl}}(r){}^4\\
 -4 f^3 l (l+1) r \omega _{\text{pl}}(r) \left(f^2 l (l+1) r \omega _{\text{pl}}''(r)+2 f l (l+1) \omega _{\text{pl}}'(r)\right)-20 f^4 r^6 \omega _{\text{pl}}(r){}^8\Big]\\
 -2 f^2 \omega _{\text{pl}}(r){}^4 \left(6 f^4 l (l+1) r^4 \omega _{\text{pl}}'(r){}^2-2 f^3 l^2 (l+1)^2 \left(-f+l^2+l+1\right)\right)\\
 +\omega ^6 \left(-48 f^2 l (l+1) r^4 \omega _{\text{pl}}(r){}^2-40 f^2 r^6 \omega _{\text{pl}}(r){}^4+2 f l (l+1) r^2 (f (15 f-6 l (l+1)-10)+1)\right)\\
 +4 f^5 l (l+1) r^3 \omega _{\text{pl}}(r){}^5 \left((1-5 f) \omega _{\text{pl}}'(r)+f r \omega _{\text{pl}}''(r)\right) +12 f^5 l (l+1) r^4 \omega _{\text{pl}}(r){}^8\\ 
+4 f^4 l (l+1) r \omega _{\text{pl}}(r){}^3 \left(f^2 l (l+1) r \omega _{\text{pl}}''(r)+(f+1) f l (l+1) \omega _{\text{pl}}'(r)\right) -4 r^6 \omega ^{10}\\ -4 f^5 l (l+1) r^2 (4 f-3 l (l+1)-1) \omega _{\text{pl}}(r){}^6
 +4 f^5 r^6 \omega _{\text{pl}}(r){}^{10}+\omega ^8 \left(12 f l (l+1) r^4+20 f r^6 \omega _{\text{pl}}(r){}^2\right)
 \end{multline}
\end{widetext}
In this case the asymptotic behavior reads $V\to-\omega^2$ as $r\to2M$ and $V\to \omega_{\rm pl}^2(r\to\infty)-\omega^2$ as $r\to\infty$. 
Obviously, the effective potential in Eq.~\eqref{eq_cfunction2} reduces to Eq.~\eqref{eq_cfunction} when $\omega_{\rm pl}={\rm const}$.

%----------------------------------------------------------------------------------------------------
\bibliographystyle{utphys}
\bibliography{Ref}

\providecommand{\href}[2]{#2}\begingroup\raggedright\begin{thebibliography}{10}

\bibitem{Barack:2018yly}
L.~Barack {\em et~al.}, ``{Black holes, gravitational waves and fundamental
  physics: a roadmap},'' \href{http://dx.doi.org/10.1088/1361-6382/ab0587}{{\em
  Class. Quant. Grav.} {\bfseries 36} no.~14, (2019) 143001},
  \href{http://arxiv.org/abs/1806.05195}{{\ttfamily arXiv:1806.05195 [gr-qc]}}.

\bibitem{1973A&A....24..337S}
N.~I. {Shakura} and R.~A. {Sunyaev}, ``{Reprint of 1973A\&A....24..337S. Black
  holes in binary systems. Observational appearance.},'' {\em \aap} {\bfseries
  500} (June, 1973) 33--51.

\bibitem{Novikov:1973kta}
I.~D. Novikov and K.~S. Thorne, ``{Astrophysics and black holes},'' in {\em
  {Proceedings, Ecole d'Eté de Physique Théorique: Les Astres Occlus}},
  pp.~343--550.
\newblock
1973.
\newblock
%%CITATION = INSPIRE-1361968;%%.

\bibitem{Sitenko:1967}
A.~G. Sitenko, {\em {Electromagnetic Fluctuations in Plasma}}.
\newblock Academic, New York, 1976.

\bibitem{Rosa:2011my}
J.~G. Rosa and S.~R. Dolan, ``{Massive vector fields on the Schwarzschild
  spacetime: quasi-normal modes and bound states},''
  \href{http://dx.doi.org/10.1103/PhysRevD.85.044043}{{\em Phys. Rev. D}
  {\bfseries 85} (2012) 044043},
  \href{http://arxiv.org/abs/1110.4494}{{\ttfamily arXiv:1110.4494 [hep-th]}}.

\bibitem{Pani:2012vp}
P.~Pani, V.~Cardoso, L.~Gualtieri, E.~Berti, and A.~Ishibashi, ``{Black hole
  bombs and photon mass bounds},''
  \href{http://dx.doi.org/10.1103/PhysRevLett.109.131102}{{\em Phys. Rev.
  Lett.} {\bfseries 109} (2012) 131102},
\href{http://arxiv.org/abs/1209.0465}{{\ttfamily arXiv:1209.0465 [gr-qc]}}.
%%CITATION = ARXIV:1209.0465;%%.

\bibitem{Pani:2012bp}
P.~Pani, V.~Cardoso, L.~Gualtieri, E.~Berti, and A.~Ishibashi, ``{Perturbations
  of slowly rotating black holes: massive vector fields in the Kerr metric},''
  \href{http://dx.doi.org/10.1103/PhysRevD.86.104017}{{\em Phys. Rev. D}
  {\bfseries 86} (2012) 104017},
  \href{http://arxiv.org/abs/1209.0773}{{\ttfamily arXiv:1209.0773 [gr-qc]}}.

\bibitem{Baryakhtar:2017ngi}
M.~Baryakhtar, R.~Lasenby, and M.~Teo, ``{Black Hole Superradiance Signatures
  of Ultralight Vectors},''
  \href{http://dx.doi.org/10.1103/PhysRevD.96.035019}{{\em Phys. Rev.}
  {\bfseries D96} no.~3, (2017) 035019},
\href{http://arxiv.org/abs/1704.05081}{{\ttfamily arXiv:1704.05081 [hep-ph]}}.
%%CITATION = ARXIV:1704.05081;%%.

\bibitem{Cardoso:2018tly}
V.~Cardoso, O.~J. Dias, G.~S. Hartnett, M.~Middleton, P.~Pani, and J.~E.
  Santos, ``{Constraining the mass of dark photons and axion-like particles
  through black-hole superradiance},''
  \href{http://dx.doi.org/10.1088/1475-7516/2018/03/043}{{\em JCAP} {\bfseries
  03} (2018) 043}, \href{http://arxiv.org/abs/1801.01420}{{\ttfamily
  arXiv:1801.01420 [gr-qc]}}.

\bibitem{Frolov:2018ezx}
V.~P. Frolov, P.~Krtous, D.~Kubiznak, and J.~E. Santos, ``{Massive Vector
  Fields in Rotating Black-Hole Spacetimes: Separability and Quasinormal
  Modes},'' \href{http://dx.doi.org/10.1103/PhysRevLett.120.231103}{{\em Phys.
  Rev. Lett.} {\bfseries 120} (2018) 231103},
\href{http://arxiv.org/abs/1804.00030}{{\ttfamily arXiv:1804.00030 [hep-th]}}.
%%CITATION = ARXIV:1804.00030;%%.

\bibitem{Dolan:2018dqv}
S.~R. Dolan, ``{Instability of the Proca field on Kerr spacetime},''
  \href{http://dx.doi.org/10.1103/PhysRevD.98.104006}{{\em Phys. Rev.}
  {\bfseries D98} no.~10, (2018) 104006},
\href{http://arxiv.org/abs/1806.01604}{{\ttfamily arXiv:1806.01604 [gr-qc]}}.
%%CITATION = ARXIV:1806.01604;%%.

\bibitem{Baumann:2019eav}
D.~Baumann, H.~S. Chia, J.~Stout, and L.~ter Haar, ``{The Spectra of
  Gravitational Atoms},''
  \href{http://dx.doi.org/10.1088/1475-7516/2019/12/006}{{\em JCAP} {\bfseries
  1912} no.~12, (2019) 006},
\href{http://arxiv.org/abs/1908.10370}{{\ttfamily arXiv:1908.10370 [gr-qc]}}.
%%CITATION = ARXIV:1908.10370;%%.

\bibitem{Brito:2015oca}
R.~Brito, V.~Cardoso, and P.~Pani,
  \href{http://dx.doi.org/10.1007/978-3-319-19000-6}{{\em {Superradiance}: {New
  Frontiers in Black Hole Physics}}}, vol.~906.
\newblock Springer, 2015.
\newblock \href{http://arxiv.org/abs/1501.06570}{{\ttfamily arXiv:1501.06570
  [gr-qc]}}.

\bibitem{Pani:2013hpa}
P.~Pani and A.~Loeb, ``{Constraining Primordial Black-Hole Bombs through
  Spectral Distortions of the Cosmic Microwave Background},''
  \href{http://dx.doi.org/10.1103/PhysRevD.88.041301}{{\em Phys. Rev. D}
  {\bfseries 88} (2013) 041301},
  \href{http://arxiv.org/abs/1307.5176}{{\ttfamily arXiv:1307.5176
  [astro-ph.CO]}}.

\bibitem{Blas:2020nbs}
D.~Blas and S.~J. Witte, ``{Imprints of Axion Superradiance in the CMB},''
  \href{http://arxiv.org/abs/2009.10074}{{\ttfamily arXiv:2009.10074
  [astro-ph.CO]}}.

\bibitem{Conlon:2017hhi}
J.~P. Conlon and C.~A. Herdeiro, ``{Can black hole superradiance be induced by
  galactic plasmas?},''
  \href{http://dx.doi.org/10.1016/j.physletb.2018.02.073}{{\em Phys. Lett. B}
  {\bfseries 780} (2018) 169--173},
  \href{http://arxiv.org/abs/1701.02034}{{\ttfamily arXiv:1701.02034
  [astro-ph.HE]}}.

\bibitem{Cardoso:2020nst}
V.~Cardoso, W.-d. Guo, C.~F. Macedo, and P.~Pani, ``{The tune of the universe:
  the role of plasma in tests of strong-field gravity},''
  \href{http://arxiv.org/abs/2009.07287}{{\ttfamily arXiv:2009.07287 [gr-qc]}}.

\bibitem{Blas:2020kaa}
D.~Blas and S.~J. Witte, ``{Quenching Mechanisms of Photon Superradiance},''
  \href{http://arxiv.org/abs/2009.10075}{{\ttfamily arXiv:2009.10075
  [hep-ph]}}.

\bibitem{Abramowicz:2011xu}
M.~A. Abramowicz and P.~Fragile, ``{Foundations of Black Hole Accretion Disk
  Theory},'' \href{http://dx.doi.org/10.12942/lrr-2013-1}{{\em Living Rev.
  Rel.} {\bfseries 16} (2013) 1},
  \href{http://arxiv.org/abs/1104.5499}{{\ttfamily arXiv:1104.5499
  [astro-ph.HE]}}.

\bibitem{Yuan:2014gma}
F.~Yuan and R.~Narayan, ``{Hot Accretion Flows Around Black Holes},''
  \href{http://dx.doi.org/10.1146/annurev-astro-082812-141003}{{\em Ann. Rev.
  Astron. Astrophys.} {\bfseries 52} (2014) 529--588},
  \href{http://arxiv.org/abs/1401.0586}{{\ttfamily arXiv:1401.0586
  [astro-ph.HE]}}.

\bibitem{2017mcp..book.....T}
K.~S. {Thorne} and R.~D. {Blandford}, {\em {Modern Classical Physics: Optics,
  Fluids, Plasmas, Elasticity, Relativity, and Statistical Physics}}.
\newblock 2017.

\bibitem{PhysRevD.45.525}
R.~Kulsrud and A.~Loeb, ``Dynamics and gravitational interaction of waves in
  nonuniform media,'' \href{http://dx.doi.org/10.1103/PhysRevD.45.525}{{\em
  Phys. Rev. D} {\bfseries 45} (Jan, 1992) 525--531}.
  \url{https://link.aps.org/doi/10.1103/PhysRevD.45.525}.

\bibitem{PhysRevE.62.2989}
J.~T. Mendon\ifmmode~\mbox{\c{c}}\else \c{c}\fi{}a, A.~M. Martins, and
  A.~Guerreiro, ``Field quantization in a plasma: Photon mass and charge,''
  \href{http://dx.doi.org/10.1103/PhysRevE.62.2989}{{\em Phys. Rev. E}
  {\bfseries 62} (Aug, 2000) 2989--2991}.
  \url{https://link.aps.org/doi/10.1103/PhysRevE.62.2989}.

\bibitem{1974ApJ...191..499P}
D.~N. {Page} and K.~S. {Thorne}, ``{Disk-Accretion onto a Black Hole.
  Time-Averaged Structure of Accretion Disk},''
  \href{http://dx.doi.org/10.1086/152990}{{\em \apj} {\bfseries 191} (July,
  1974) 499--506}.

\bibitem{Arvanitaki:2010sy}
A.~Arvanitaki and S.~Dubovsky, ``{Exploring the String Axiverse with Precision
  Black Hole Physics},''
  \href{http://dx.doi.org/10.1103/PhysRevD.83.044026}{{\em Phys.Rev.}
  {\bfseries D83} (2011) 044026},
\href{http://arxiv.org/abs/1004.3558}{{\ttfamily arXiv:1004.3558 [hep-th]}}.
%%CITATION = ARXIV:1004.3558;%%.

\bibitem{Brito:2014wla}
R.~Brito, V.~Cardoso, and P.~Pani, ``{Black holes as particle detectors:
  evolution of superradiant instabilities},''
  \href{http://dx.doi.org/10.1088/0264-9381/32/13/134001}{{\em Class. Quant.
  Grav.} {\bfseries 32} no.~13, (2015) 134001},
\href{http://arxiv.org/abs/1411.0686}{{\ttfamily arXiv:1411.0686 [gr-qc]}}.
%%CITATION = ARXIV:1411.0686;%%.

\bibitem{1981A&A....96..293B}
R.~A. {Breuer} and J.~{Ehlers}, ``{Propagation of electromagnetic waves through
  magnetized plasmas in arbitrary gravitational fields},'' {\em \aap}
  {\bfseries 96} no.~1-2, (Mar., 1981) 293--295.

\bibitem{Ellis:1971pg}
G.~Ellis, ``{Relativistic cosmology},''
  \href{http://dx.doi.org/10.1007/s10714-009-0760-7}{{\em Gen. Rel. Grav.}
  {\bfseries 41} (2009) 581--660}.

\bibitem{Raffelt:1996wa}
G.~Raffelt, {\em {Stars as laboratories for fundamental physics}: {The
  astrophysics of neutrinos, axions, and other weakly interacting particles}}.
\newblock 5, 1996.

\bibitem{Galtsov:1984ixy}
D.~Gal'tsov, G.~Pomerantseva, and G.~Chizhov, ``{Behavior of massive vector
  particles in a Schwarzschild field},''
  \href{http://dx.doi.org/10.1007/BF00893117}{{\em Sov. Phys. J.} {\bfseries
  27} (1984) 697--700}.

\bibitem{Detweiler:1980uk}
S.~L. Detweiler, ``{Klein-Gordon equation and rotating black holes},''
  \href{http://dx.doi.org/10.1103/PhysRevD.22.2323}{{\em Phys. Rev. D}
  {\bfseries 22} (1980) 2323--2326}.

\bibitem{Dolan:2007mj}
S.~R. Dolan, ``{Instability of the massive Klein-Gordon field on the Kerr
  spacetime},'' \href{http://dx.doi.org/10.1103/PhysRevD.76.084001}{{\em
  Phys.Rev.} {\bfseries D76} (2007) 084001},
\href{http://arxiv.org/abs/0705.2880}{{\ttfamily arXiv:0705.2880 [gr-qc]}}.
%%CITATION = ARXIV:0705.2880;%%.

\bibitem{Brito:2013wya}
R.~Brito, V.~Cardoso, and P.~Pani, ``{Massive spin-2 fields on black hole
  spacetimes: Instability of the Schwarzschild and Kerr solutions and bounds on
  the graviton mass},''
  \href{http://dx.doi.org/10.1103/PhysRevD.88.023514}{{\em Phys. Rev.}
  {\bfseries D88} (2013) 023514},
\href{http://arxiv.org/abs/1304.6725}{{\ttfamily arXiv:1304.6725 [gr-qc]}}.
%%CITATION = ARXIV:1304.6725;%%.

\bibitem{Brito:2020lup}
R.~Brito, S.~Grillo, and P.~Pani, ``{Black Hole Superradiant Instability from
  Ultralight Spin-2 Fields},''
  \href{http://dx.doi.org/10.1103/PhysRevLett.124.211101}{{\em Phys. Rev.
  Lett.} {\bfseries 124} no.~21, (2020) 211101},
  \href{http://arxiv.org/abs/2002.04055}{{\ttfamily arXiv:2002.04055 [gr-qc]}}.

\bibitem{1952MNRAS.112..195B}
H.~{Bondi}, ``{On spherically symmetrical accretion},''
  \href{http://dx.doi.org/10.1093/mnras/112.2.195}{{\em \mnras} {\bfseries 112}
  (Jan., 1952) 195}.

\bibitem{Ferrari:2007rc}
V.~Ferrari, L.~Gualtieri, and S.~Marassi, ``{A New approach to the study of
  quasi-normal modes of rotating stars},''
  \href{http://dx.doi.org/10.1103/PhysRevD.76.104033}{{\em Phys. Rev. D}
  {\bfseries 76} (2007) 104033},
  \href{http://arxiv.org/abs/0709.2925}{{\ttfamily arXiv:0709.2925 [gr-qc]}}.

\bibitem{Dima:2020rzg}
A.~Dima and E.~Barausse, ``{Numerical investigation of plasma-driven
  superradiant instabilities},''
  \href{http://dx.doi.org/10.1088/1361-6382/ab9ce0}{{\em Class. Quant. Grav.}
  {\bfseries 37} no.~17, (2020) 175006},
  \href{http://arxiv.org/abs/2001.11484}{{\ttfamily arXiv:2001.11484 [gr-qc]}}.

\bibitem{Kojima:1992ie}
Y.~Kojima, ``{Equations governing the nonradial oscillations of a slowly
  rotating relativistic star},''
  \href{http://dx.doi.org/10.1103/PhysRevD.46.4289}{{\em Phys. Rev. D}
  {\bfseries 46} (1992) 4289--4303}.

\bibitem{Pani:2013pma}
P.~Pani, ``{Advanced Methods in Black-Hole Perturbation Theory},''
  \href{http://dx.doi.org/10.1142/S0217751X13400186}{{\em Int. J. Mod. Phys.}
  {\bfseries A28} (2013) 1340018},
\href{http://arxiv.org/abs/1305.6759}{{\ttfamily arXiv:1305.6759 [gr-qc]}}.
%%CITATION = ARXIV:1305.6759;%%.

\bibitem{Cardoso:2013opa}
V.~Cardoso, I.~P. Carucci, P.~Pani, and T.~P. Sotiriou, ``{Matter around Kerr
  black holes in scalar-tensor theories: scalarization and superradiant
  instability},'' \href{http://dx.doi.org/10.1103/PhysRevD.88.044056}{{\em
  Phys. Rev. D} {\bfseries 88} (2013) 044056},
  \href{http://arxiv.org/abs/1305.6936}{{\ttfamily arXiv:1305.6936 [gr-qc]}}.

\bibitem{Cardoso:2013fwa}
V.~Cardoso, I.~P. Carucci, P.~Pani, and T.~P. Sotiriou, ``{Black holes with
  surrounding matter in scalar-tensor theories},''
  \href{http://dx.doi.org/10.1103/PhysRevLett.111.111101}{{\em Phys. Rev.
  Lett.} {\bfseries 111} (2013) 111101},
  \href{http://arxiv.org/abs/1308.6587}{{\ttfamily arXiv:1308.6587 [gr-qc]}}.

\bibitem{Dolan:2012yt}
S.~R. Dolan, ``{Superradiant instabilities of rotating black holes in the time
  domain},'' \href{http://dx.doi.org/10.1103/PhysRevD.87.124026}{{\em
  Phys.Rev.} {\bfseries D87} (2013) 124026},
\href{http://arxiv.org/abs/1212.1477}{{\ttfamily arXiv:1212.1477 [gr-qc]}}.
%%CITATION = ARXIV:1212.1477;%%.

\end{thebibliography}\endgroup
%----------------------------------------------------------------------------------------------------

\end{document}